\documentclass[preprint,pre,aps,superscriptaddress,a4paper]{revtex4}
\usepackage[dvips]{graphicx}
\usepackage[T1]{fontenc}
\usepackage{epsfig,amsmath,amsfonts,amssymb,
color,
}

\begin{document}

\title{
Effect of antagonistic salt on confined near-critical mixture
}
\author{Faezeh Pousaneh}
\affiliation{Institute of Physical Chemistry, Polish Academy of Sciences,
  Kasprzaka 44/52, PL-01-224 Warsaw, Poland}
\author{Alina Ciach}
\affiliation{Institute of Physical Chemistry, Polish Academy of Sciences,
  Kasprzaka 44/52, PL-01-224 Warsaw, Poland}
\begin{abstract}
 We consider a near-critical binary mixture with addition of antagonistic salt 
 confined between weakly charged and selective surfaces. A mesoscopic functional for this system is developed 
 from a microscopic description by a systematic coarse-graining procedure. The functional reduces 
 to the Landau-Brazovskii
 functional for amphiphilic systems for sufficiently large ratio between the correlation length in the critical 
 binary mixture  and the screening length. Our theoretical result agrees with the experimental observation 
 [Sadakane et.al. J. Chem. Phys. {\bf 139}, 234905 (2013)] that the antagonistic salt and surfactant both lead to
 a similar mesoscopic structure. 
 For very small salt concentration $\rho_{ion}$ the Casimir  potential 
 is the same as in a presence  of inorganic salt. For larger $\rho_{ion}$ the Casimir potential 
 takes a minimum followed by a maximum for separations of order of tens of nanometers, and   exhibits an
 oscillatory decay very close to the critical 
 point. For separations of  tens of nanometers the potential between surfaces with a linear 
 size of  hundreds of nanometers can be of order of $k_BT$. We have verified that in the experimentally
 studied samples [Sadakane et.al. J. Chem. Phys. {\bf 139}, 234905 (2013), Leys et.al. Soft Matter {\bf 9}, 9326 (2013)]
 the decay length is too small compared to the period of oscillations of the Casimir potential,
 but the oscillatory force could be observed closer to the critical point.
 
\end{abstract}
\maketitle
\section{Introduction}
Addition of small amount of salt to a binary mixture near a demixing critical point  can 
significantly change its properties. When  inorganic salt is added to water and organic liquid, the two-phase region
enlarges~\cite{fuess:99:0}. In contrast, antagonistic (or amphiphilic) salt added to such mixture leads to shrinking 
of the two phase region;
it can even disappear when the amount of salt is large enough 
\cite{sadakane:13:0}. 
In addition, in the case of the antagonistic salt
a peak in the structure factor for
the wavenumber $k>0$ was observed in the one-phase region~\cite{sadakane:09:0,sadakane:11:0,leys:13:0,sadakane:13:0}.
The peak indicates thermodynamically stable inhomogeneities on 
the length scale $\sim 10nm$~\cite{sadakane:13:0,leys:13:0}.
Moreover, in a few cases a lamellar phase was observed for some region of the phase diagram
\cite{sadakane:09:0,sadakane:13:0,leys:13:0}. For low salt concentration
the shape of the structure factor was described with a good accuracy by the formula obtained by Onuki and Kitamura
\cite{onuki:04:0}. For larger amount of salt (when the two-phase region disappears)
the experimental structure factor was described with better accuracy by the formula derived earlier 
for bicontinuous microemulsion or sponge phases~\cite{sadakane:13:0,lei:97:0,porcar:03:0}. In addition, the structure factor of the lamellar 
phase was fitted with a good accuracy~\cite{sadakane:13:0} by the formula developed for a stack of
membranes \cite{sadakane:13:0,nallet:93:0}. These observations strongly suggest that the key features of  the mesoscopic structure do not depend on 
whether  antagonistic salt or 
surfactant is added to a mixture of inorganic and organic solvents. In this work we address the question of the above
similarity on a theoretical level by comparing the Landau-type functionals for the two systems. 
For this purpose we first develop a 
Landau-type functional starting from a microscopic density functional theory and using the same coarse-graining
procedure as used earlier for the hydrophilic salt
~\cite{ciach:10:0,pousaneh:14:0}. In the next step we verify under what conditions the Brazovskii functional
 of the solvent concentration, developed earlier for
 amphiphilic systems~\cite{gompper:94:0,ciach:01:2,brazovskii:75:0} can be obtained.

The effects of confinement on the near-critical mixture with antagonistic salt were not investigated experimentally 
yet.
Theoretical investigations based on the theory of Onuki and Kitamura~\cite{onuki:04:0} focused very briefly on  colloid 
particles immersed
in the critical mixture with salt~\cite{okamoto:11:0}. In contrast, the effective potentials between
a flat substrate and a colloid particle
immersed in the critical mixture with hydrophilic salt, 
and between two particles
in such mixture were  measured in several 
impressive experiments \cite{hertlein:08:0,nellen:11:0,gambassi:09:0,bonn:09:0,nguyen:13:0}. 
Similarity between the structure factors in the investigated system and sponge phases suggests 
that an oscillatory force between confining surfaces, observed 
 experimentally for surfactant mixtures \cite{antelmi:99:0}, could occur in a presence of antagonistic salt too.

The effective potential between surfaces confining the critical mixture with ions is interesting from both 
the fundamental and the practical point of view. Surfaces with similar adsorption preferences attract each other,
while surfaces with different adsorption preferences repel each other~\cite{gambassi:09:0, vasiliev:11:0, krech:96:0, krech:94:0} when the confined fluid is 
near its critical point. 
The range of this so called thermodynamic Casimir
potential \cite{krech:94:0,gambassi:09:0} is equal to the bulk correlation length 
$\xi\propto |T-T_c|^{-\nu}$ with $\nu\approx 0.63$, and becomes
macroscopic when the critical temperature $T_c$ is approached. When the surfaces are charged, then depending on the
surface charges electrostatic repulsion or attraction is added to the Casimir potential. The resulting potential can have 
a minimum or a maximum, depending on the ratio between $\xi$ and the screening length $\lambda_D$. 
By changing the salt concentration and the temperature, one can change $\lambda_D$ and $\xi$, and tune the shape of the
potential between the surfaces, e.g. the surfaces of particles \cite{hertlein:08:0,nellen:11:0,gambassi:09:0,nguyen:13:0}. Temperature can be changed in a reversible manner, 
and reversible structural changes can be
induced \cite{bonn:09:0, nguyen:13:0, dang:13:0}. 
In particular, when the effective potential has a minimum, then analogs of the gas-liquid and liquid-solid 
transitions between the particles were observed~\cite{nguyen:13:0}. 

The experimental results for inorganic salt show that
the effective potential
cannot be described by just a sum of the Casimir and the electrostatic potential \cite{gambassi:09:0, pousaneh:14:1}. 
In particular,
attraction was measured for some temperature range between 
like charge hydrophilic and hydrophobic surfaces~\cite{nellen:11:0}, whereas
both the Casimir and the electrostatic potentials are repulsive in this case. Both potentials, however, 
 can be modified because of different solubility of the ions in the two components of the solvent. 
 For this reason quite different behavior of the effective potential between charged and
 selective surfaces can be expected for the 
 antagonistic salt~\cite{okamoto:11:0}, and we address this question here. 
 
 Several groups tried to explain the experiment in Ref.~\cite{nellen:11:0}. The attraction for a range of 
 temperatures was 
 obtained in two different approaches. The first one is based on the theory of Onuki and Kitamura~\cite{onuki:04:0}.
 The attraction appears when the solubility in water  of the anion and the cation  differ significantly from each other
  and the
 hydrophobic surface is neutral~\cite{bier:11:0,samin:12:0}. On the other hand, in the theory developed in 
 Ref.\cite{ciach:10:0} for hydrophilic ions,  the concentration profile near the charged hydrophobic surface can be
 nonmonotonic, because the attraction of the organic particles competes with the attraction of the
 hydrated ions~\cite{pousaneh:11:0,pousaneh:14:0}.
 When $\lambda_D>\xi$, excess of water may appear at some distance from the surface, therefore it may act
 as a hydrophilic one. 
 The shapes of the effective potential
 for different temperatures in this theory are quite similar to the experimental curves,
 but the fitting was not attempted yet~\cite{pousaneh:14:1}.
 Importantly, good quantitative agreement
 between this theory and the experiments conducted for $\lambda_D<\xi$ ~\cite{hertlein:08:0,gambassi:09:0}
 was obtained in Ref.\cite{pousaneh:12:0}, 
 in contrast to the theories based on the Onuki and Kitamura functional. 
 In this work we develop a Landau-type theory for
 the antagonistic salt following the same strategy as in Ref.\cite{ciach:10:0,pousaneh:14:0}.
 
 In the Onuki and Kitamura model the critical binary mixture is described by the phenomoenological 
 Landau functional of the concentration $\phi$.
 The ions are treated as a two-component ideal gas whose particles interact with the Coulomb potential, and the 
 corresponding entropy and electrostatic energy are added to the Landau functional. Finally, the coupling between the
 critical mixture and the ions of the form $\phi(w_+n_++w_-n_-)$ is added, where $n_i$ and $w_i$ denote the 
 density and the preferential salvation of the $i$-th ion. Note that in this model the van der Waals 
 interactions between 
 the ions are neglected, and the entropy of the four-component mixture is approximated by a sum of 
 the entropy of the two
 two-component subsystems. For this reason the coupling between the two subsystems beyond the above bilinear term
 is not taken into account. 
 
 The full microscopic density-functional theory of a four-component mixture is more accurate and justified
 than a phenomenological model,
 but in practice it is too difficult. Moreover, such a detailed description is not necessary when the
 characteristic length scales are mesoscopic. On the other hand, a microscopic theory is a good starting point
 for a derivation
 of the Landau-type  functional by a coarse-graining procedure. 
 Such a strategy was successful in the case of the hydrophilic salt \cite{pousaneh:14:1, ciach:10:0}.
 In the next section we develop in the same way a Landau-type functional for the 
 antagonistic salt added to the near-critical binary solvent. In sec.3 we focus on the disordered phase in the bulk.
  We calculate the correlation function, 
  and compare our functional with the Landau-Brazovskii functional for microemulsions. 
  In sec.4 we consider the concentration
 and charge profiles in a semiinfinite system and in a  slit, and calculate  
 the effective potential between the surfaces. In sec.5 a
  comparison of the structure factor with the experimental results reported in Ref.\cite{sadakane:13:0,leys:13:0} 
  is performed. For the model parameters obtained from the fitting  of the structure factor to the experimental results
  the concentration profiles  and the effective potential are computed. Sec.6
  contains summary and discussion. In particular, we compare the effective potential obtained in the models
  for the hydrophilic and the antagonistic salt.

\section{The model}
\label{generic}
We consider a four-component mixture containing water, organic liquid and  antagonistic salt 
which has hydrophilic cations and hydrophobic  anions. 
The fluid is in contact with a
reservoir with fixed temperature and chemical potential of all
the components. In equilibrium the distribution of the
components corresponds to the minimum of the
grand potential 
\begin{eqnarray}
 \label{OmegaL}
\Omega=  -PV \;\;\;\;\;\;\;\;\;\; \;\;\;\;\;\;\;\;\;\; \;\;\;\;\;\;\;\;\;\; \;\;\;\;\;\;\;\;\;\; \;\;\;\;\;\;\;\;\;\; \;\;\;\;\;\;\;\;\;\;
\\
\nonumber
 = U_{vdW}+U_{el}-TS-\int_Vd{\bf r}^*\mu_i\rho_i^*({\bf r}),\;\;\;\;\;\;\;\;\;\; \;\;\;\;\;\;\;
\end{eqnarray}
where $P$, $V$ and  $S$ are pressure, volume and entropy respectively, and   $U_{el}$, $U_{vdW}$ are the   
electrostatic and the van der Waals contributions to the internal energy. $\rho_i ({\bf r})$  
and $\mu _i$ are the  local
number density 
and the chemical 
potential of the $i$-th component respectively, with $i=w,l$ corresponding 
to water and organic solvent (for example lutidine or methylpyridine), 
and $i=+,-$ corresponding to the cations and the anions. 
We  consider dimensionless distance $r^*=r/a$ and dimensionless densities,
i.e. the length is measured in units of $a$, 
and $\rho_i^*=a^3 N_i/V$, where $a^3$ is the average volume per particle in the liquid phase, and $N_i$ denotes
 the number of the  $i$-th kind particles in the volume $V$.
 We neglect compressibility of the liquid and assume 
that the total density is fixed, 
\begin{equation}
\label{fixrho}
 \sum_{i=\{w,l,+,-\}}\rho_i^*=1.
\end{equation}
 The microscopic details, in particular different sizes of molecules are disregarded, 
since we are interested in the local densities 
in the regions much larger than $a$.  
To simplify the notation we shall omit the asterisk for the dimensionless distance.

The electrostatic energy  is given by \cite{barrat:03:0} 
\begin{equation}
\label{el}
U_{el}=\int_V d{\bf r} \left[ -\frac{\epsilon}{8\pi}(\bigtriangledown{\psi_{el}({\bf r})})^2
+e{\rho_q^*({\bf r})} {\psi_{el}({\bf r})} \right],\\
 \end{equation} 
where ${\psi_{el}({\bf r})}$ is the electrostatic potential which satisfies the Poisson equation (note that the 
length $|{\bf r}|$ and density are dimensionless)
\begin{eqnarray}
\label{Poisson}
 \bigtriangledown ^2 \psi_{el} ({\bf r})=-\frac {4\pi e} {a\epsilon}\rho_q^*({\bf r}).
\end{eqnarray}
 We neglect the dependence of  the dielectric constant $\epsilon$ on the solvent 
concentration (\cite{ciach:10:0,pousaneh:14:0}). $\rho_q^*({\bf r})$ is the local dimensionless charge density 
\begin{eqnarray}
\label{rhoqdef}
 \rho_q^*({\bf r})=\rho_+^*({\bf r})-\rho_-^*({\bf r}),
\end{eqnarray}
 and $e$ is the elementary charge. \\
 
We assume the usual form of the internal energy ${\cal U}_{vdW}$,
\begin{eqnarray}
\label{vW}
{\cal U}_{vdW}
=\int_{V}d{\bf r}\int_{V} d{\bf r}' \frac{1}{2}\rho_i^*({\bf r})V_{ij}({\bf r}-{\bf r}') g_{ij}({\bf r}-{\bf r}')
\rho_j^*({\bf r}'),
\end{eqnarray}
where $V_{ij}$ and $g_{ij}$ are the vdW
interaction and the pair correlation function between the corresponding components respectively. 
The very complex expression for ${\cal U}_{vdW}$ can be approximated by a much simpler form  in  systems with particular
properties of $V_{ij}$,
and in special thermodynamic states, such as the neighborhood of the critical point.
Close to the  critical point and for small amount of ions the correlation and screening  lengths are
large, and the local densities
vary slowly on the microscopic length scale. Thus, the density  $\rho_j^*({\bf r'})$ can be Taylor expanded 
about ${\bf r}$. In the case of attractive interactions the expansion can be truncated at the
 second-order therm ~\cite{ciach:10:0,ciach:13:03}.
 As a result, ${\cal U}_{vdW}$ is given by a single
integral over ${\bf r}$ with  the integrand
depending on the densities and their gradients at ${\bf r}$, and on the zeroth, 
$J_{ij}=-\int_{V}d{\bf r}V_{ij}({\bf r})g_{ij}({\bf r})$,  and the second, 
$\bar J_{ij}=-\int_{V}d{\bf r}V_{ij}({\bf r})g_{ij}({\bf r})r^2/6$,
moments of 
$V_{ij}g_{ij}$~\cite{ciach:10:0,ciach:13:03}.
Since we consider two inorganic (water and cation) and two organic (metylpyridine and anion) components,
we can further simplify the expression for  $U_{vdW}$ by assuming appropriate relations between $J_{ij}$.   
We assume that the differences between the interactions of all 
the inorganic components are negligible; likewise, we neglect the differences  between the interactions of all 
the organic components. Accordingly, we assume

\begin{eqnarray}
\label{J1}
J_{ww}\approx J_{++} \approx J_{w+}\\\nonumber
J_{ll} \approx J_{--} \approx J_{l-}\\\nonumber
J_{wl}\approx J_{w-} \approx J_{l+} \approx J_{+-}.
\end{eqnarray}
with analogous assumptions for $\bar J_{ij}$.
After some algebra we obtain
\begin{eqnarray}
\label{uvdW0}
{\cal U}_{vdW}=U_{vdW}[\Phi]
+J_1\int_V  \Phi({\bf r})d{\bf r} +J_0V ,
\end{eqnarray}
where
\begin{eqnarray}
\label{uvdW}
U_{vdW}[\Phi]=\frac{J}{2}\int_V \Big[-b{\Phi({\bf r})}^2 + (\nabla {\Phi({\bf r})})^2 \Big]d{\bf r}
\end{eqnarray}
and
\begin{eqnarray}
\label{OP}
\Phi({\bf r})= \rho_w^*({\bf r})-\rho_l^*({\bf r}) +\rho_q^*({\bf r})
\end{eqnarray}
is the difference between the local dimensionless density of the inorganic and organic components (see (\ref{rhoqdef})),
$J_1=\frac{1}{4}(J_{ww}-J_{ll})$, $J_0=\frac{1}{8}(J_{ww}+J_{ll}+2J_{wl})$, 
 and 
 \begin{eqnarray}
 J=\frac{1}{4}(\bar J_{ww}+\bar J_{ll}-2\bar J_{wl})
 \end{eqnarray}
 is the energy parameter relevant for the phase separation. 
 The parameter $b$
 is associated with the difference between the second and the zeroth moments of the interactions. 
 In the case of the lattice model with nearest-neighbor interactions $b=2d$ in a $d$-dimensional system.
 We neglect the constant term in (\ref{uvdW0}), 
 and add the linear term in (\ref{uvdW0}) to $-\mu_i\sum_i\rho_i^*({\bf r})$ in Eq.(\ref{OmegaL}). As a consequence,  
  the internal energy is given in Eq.(\ref{uvdW}) and the  chemical potentials are modified.

Finally, 
we postulate the lattice gas or ideal-mixing 
form for the entropy
\begin{equation}
\label{S}
-TS=k_BT\int_V d{\bf r}\sum_{\{i= w, l, +, -\}}\rho_i^*({\bf r})\ln\rho_i^*({\bf r}).
\end{equation}
Eq.~(\ref{S}) should be expressed in terms of the new variables $\Phi$, $\rho_q^*$ and the dimensionless  ion density
 \begin{eqnarray}
\label{rhoion}
\rho_{ion}^*({\bf r})=\rho_+^*({\bf r})+\rho_-^*({\bf r}),
\end{eqnarray}
since there are 3 independent variables when (\ref{fixrho}) holds.
 
 In this theory the thermal equilibrium for fixed temperature, total density and chemical potentials 
 corresponds to the global minimum of the grand potential (\ref{OmegaL}) 
 with the electrostatic and the vdW
 contributions to the internal energy given in Eqs.(\ref{el}) and (\ref{uvdW}) respectively,
 and the entropy given in (\ref{S}).
  The equilibrium forms of $\Phi({\bf r})$ and $\rho_{ion}^*({\bf r})$  are the  solutions of the 
Euler-Lagrange  (EL) equations obtained from the extremum conditions 
$\delta \Omega/\delta {\rho_{ion}^*({\bf r})}=0=\delta \Omega/\delta \Phi({\bf r})=0=\delta \Omega/\delta \rho_q^*({\bf r})$.
Note that neither $U_{el}$ nor $U_{vdW}$ depends on $\rho_{ion}^*({\bf r})$. For this reason $\rho_{ion}^*({\bf r})$ 
can be easily expressed 
 in terms of $\Phi({\bf r})$ and $\rho_q^*({\bf r})$  with the help of the equation 
 $\delta \Omega/\delta {\rho_{ion}^*({\bf r})}=0$ 
 (see Appendix A), and  the grand potential
becomes a functional of two fields, $\Phi({\bf r})$ and $\rho_{q}^*({\bf r})$.

\section{properties of the disordered phase}
\label{corr}
\subsection{Landau-type functional in the Gaussian approximation}
In thermodynamic conditions corresponding 
to stability of a uniform fluid $\rho_q^*({\bf r})=0$ and both $\Phi({\bf r})$ and  $\rho_{ion}^*({\bf r})$
are independent of the space position, $\Phi({\bf r})=\bar \Phi$ and $\rho_{ion}^*({\bf r})=\bar \rho_{ion}^*$.
 For the stability analysis and in order to calculate the correlation functions we introduce  
  the  Landau-Ginzburg (LG) functional 
\begin{equation}
\label{L0}
 {\cal L}[\varPhi({\bf r}), \rho_q^*({\bf r})]=\beta \Omega[ \Phi({\bf r}),\rho_{ion}^*({\bf r}),\rho_q^*({\bf r})]-
 \beta\Omega[\bar\Phi,\bar\rho_{ion}^*,0],
\end{equation}
where $\rho_{ion}^*({\bf r})$ is expressed in terms of $\Phi({\bf r})$ and $\rho_q^*({\bf r})$ (Appendix A) and
 \begin{eqnarray}
\label{fields}
 \varPhi({\bf r})=\Phi({\bf r})-\bar \Phi.
\end{eqnarray}

 We write the internal energy  in the Fourier representation (Appendix B).
  Next we Taylor-expand the entropy minimized with respect to $\rho_{ion}^*$ and approximate it by
 a second-order polynomial in $\varPhi$ and $\rho_q^*$. This way we obtain the LG functional
 in the Gaussian approximation,
\begin{equation}
\label{calL}
 {\cal L}_G[\varPhi({\bf k}),\rho({\bf k})]=\frac{1}{2}\int_V \frac{d{\bf k}}{(2\pi)^d} 
 \bigg \{ \tilde C_{\Phi\Phi}( k) \tilde \varPhi({\bf k}) 
\tilde \varPhi(-{\bf k})+2 \tilde C_{\Phi q}( k)   \tilde \varPhi({\bf k}) 
\tilde \rho_q^*(-{\bf k})
+ \tilde C_{q q}( k)   \tilde \rho_q^*({\bf k})\tilde \rho_q^*(-{\bf k})    \bigg\},
\hskip2cm
 \end{equation} 
where
\begin{equation}
\label{Ck}
 \left\{ 
  \begin{array}{l }
     \tilde C_{\Phi\Phi}( k)= \beta^* (\xi_0^{-2}+k^2), 
\vspace{0.7cm}\\
  \tilde C_{\Phi q}( k)= -\frac{1}{(1-\bar \rho_{ion}^*-\bar\Phi^2)},
\vspace{0.7cm}\\
 \tilde C_{q q}( k)=  \frac{1}{\bar \rho_{ion}^*}+\frac{1}{(1-\bar \rho_{ion}^*-\bar\Phi^2)}
 +\frac{1}{\bar \rho_{ion}^*\lambda_D^{*2}k^2  } ,
  \end{array} \right.
\end{equation}\\
with $1/\beta^*=T^*=k_BT/J$, the dimensionless screening length $\lambda_D^*$ given by
\begin{equation}
\label{lambdaD}
\lambda_D^{*2}=\lambda_D^2/a^2=\frac{a\epsilon}{4\pi e^2 \beta \bar\rho_{ion}^*},
 \end{equation} 
and with
\begin{equation}
\label{xi}
\xi_0= \bigg( \frac{T^*}{ (1-\bar \rho_{ion}^*-\bar\Phi^2)}-b\bigg)^{-1/2}.
\end{equation}

For $ \tilde \rho_q^*({\bf k})=0$ Eq.(\ref{calL}) reduces to the standard Landau functional for
the upper critical point.  $\xi_0$ is the dimensionless correlation length 
in the case of $\tilde \rho_q^*({\bf k})=0$ and $T^*>T^*_c$,  where $T^*_c=b(1-\bar \rho_{ion}^*)$ is the critical
temperature in the reduced units.
$\bar\Phi=0$ at  the critical point. The temperature of the upper critical point decreases after 
addition of the solute, because  the entropy  increases, and the energy is not changed in this model (see (\ref{J1})).
In the case of the lower critical point $T^*-T^*_c$ in (\ref{xi}) should be replaced by $|T^*-T^*_c|$.
 
 \subsection{The structure and the boundary of stability of the disordered phase}

According to the density functional theory \cite{evans:79:0} the correlation function for concentration 
fluctuations, $\tilde G( k)$ 
is given by
\begin{equation}
\label{G0}
\tilde G( k)= \frac{\tilde C_{q q }( k)}{\tilde C_{\Phi \Phi}( k)\tilde C_{q q }( k)
-\tilde C_{\Phi q}( k)^2},
\end{equation}
and its explicit form  can be easily obtained from Eq.~(\ref{Ck}),
\begin{equation}
\label{G}
\tilde G( k)=  \xi_0^{2}T^* \bigg\{ 1+\xi_0^2k^2 
-\frac{ a_N\xi_0^2 k^2}{\lambda_D^{*2} k^2 +a_D}\bigg\}^{-1}.
\end{equation}
We have introduced
\begin{eqnarray}
\label{aN}
    a_N= \frac{T^*\bar \rho_{ion}^* \lambda_D^{*2}}{ (1-\bar \rho_{ion}^*-\bar\Phi^2)(1-\bar\Phi^2)}
\end{eqnarray}
and
\begin{equation}
\label{aD}
 a_D=\frac{(1-\bar \rho_{ion}^*-\bar\Phi^2)}{(1-\bar\Phi^2)}.
\end{equation}
%
Our formula (\ref{G}) is very similar to the one obtained in Ref.\cite{onuki:04:0}. The $k$-dependence is the same, but
in Ref.\cite{onuki:04:0}  $a_D=1$, and our expression for $a_N$ is somewhat different.
We should stress that in our theory $\xi_0$ is equal to the correlation length
in the system with suppressed charge waves. 
 
  
  Note that $ {\cal L}_G[\varPhi,\rho_q]$ in Eq.(\ref{calL}) can be minimized with respect to $\rho_q^*$,
  because $\tilde C_{qq}(k)>0$. 
  At the minimum 
  $\tilde\rho_q^*({\bf k})=-\frac{\tilde C_{\Phi q}(k)}{\tilde C_{qq}(k)}\tilde \varPhi({\bf k})$
  , and the functional takes the form
   \begin{equation}
   \label{LC}
    {\cal L}_G[\varPhi]=\frac{1}{2}\int_V \frac{d{\bf k}}{(2\pi)^d}  \tilde C(k) \tilde \varPhi({\bf k}) 
\tilde \varPhi(-{\bf k})
   \end{equation}
where (see (\ref{G}))
\begin{equation}
 \tilde C(k) = \tilde G^{-1}(k).
\end{equation}
 $\tilde C(k)$ takes the minimum 
 \begin{equation}
 \label{Ck0}
 \tilde C(k_0)=\beta^*\Big[
 \big(\frac{\lambda_D^*}{\xi_0}\Big)^{2}-(\sqrt a_N-\sqrt a_D)^2\Big]\lambda_D^{*-2}
 \end{equation}
 for $k=k_0$ with
\begin{equation}
\label{k0}
k_0^2\lambda_D^{*2}=\sqrt{a_D}
(\sqrt{a_N} -\sqrt{a_D})
\end{equation}
where $a_N$ and $a_D$ are given in (\ref{aN}) and (\ref{aD}).
When $a_N<a_D$, then  $\tilde C(k)$ takes the minimum at $k=0$.
The minimum of $\tilde C(k)$ (or the maximum of $\tilde G(k)$)
with $k_0>0$ occurs only if $a_N>a_D$, or explicitly
\begin{equation}
 (k_BT)^2>J\frac{4\pi e^2}{a\epsilon} (1-\bar\rho_{ion}^*-\bar\Phi^2)^2.
\end{equation}
In our theory developed for an upper critical point the inhomogeneous
structure in the disordered phase can occur when the thermal energy is larger 
than the geometric mean of the energy cost of the salt
molecule dissociation
and the vdW energy gain of phase-separated system, $J$. At the same time $J$ is the energy 
cost of a local interface 
(see the term $\frac{J}{2}\int d{\bf r}(\nabla \Phi)^2$
in Eq.(\ref{uvdW})), and $4\pi e^2/(a\epsilon)$ is the energy gain when the two ions approach each other from
the two sides of the interface.

The correlation function in the real-space representation can be obtained by the pole analysis of
$\tilde G(k)$.
We have found that the damped oscillatory and the monotonic decay of correlations occur for 
$(\sqrt a_N-\sqrt a_D)^2<(\lambda_D^{*}/\xi_0)^2<(\sqrt a_N+\sqrt a_D)^2$, and
$(\lambda_D^{*}/\xi_0)>(\sqrt a_N+\sqrt a_D)$ respectively. The boundary between these two cases, 
$(\lambda_D^{*}/\xi_0)=(\sqrt a_N+\sqrt a_D)$ is called the disorder surface in the $(T^*,\bar\rho_{ion}^*,\bar\Phi)$
phase diagram, as in the amphiphilic systems~\cite{ciach:01:2}.

When $(\sqrt a_N-\sqrt a_D)=(\lambda_D^{*}/\xi_0)$ then  $\tilde C(k_0)=0$ (i.e. $\tilde G(k_0)\to \infty$), 
and the homogeneous phase is at the boundary of stability in the mean-field approximation (MF). 
At the corresponding $\lambda$- surface in the 
$(T^*,\bar\rho_{ion}^*,\bar\Phi)$ phase space 
(we use this name by analogy with the $\lambda$-line \cite{ciach:00:0} in 
the two-dimensional  phase space) $k_0$
is given by
\begin{equation}
 k_0^2\lambda_D^*\xi_0=\sqrt a_D\approx 1.
\end{equation}
The wavelength of the concentration wave at the$\lambda$- surface   is equal 
to a geometric mean
of the inverse correlation length and the inverse screening length. This relation allows for a quick verification
if mesoscopic inhomogeneities ($k_0\sim 2\pi/10 nm^{-1}$) can occur in the investigated experimental system. 
 \subsection{Comparison with microemulsions}
When we limit ourselves to the  most probable  concentration waves, with   the wavenumbers  
 near the maximum of the structure factor, $k\simeq k_0>0$,
then we can make the approximation
\begin{equation}
\label{C}
 \tilde C(k) \simeq \tilde C(k_0)+ c(k^2-k_0^2)^2.
\end{equation}
We took into account that $\tilde C(k)$ is a function of $k^2$, and truncated the Taylor expansion
in $k^2$ about $k_0^2$ at the second-order term.  After inserting  (\ref{C}) in (\ref{LC})
we  obtain the Landau-Brazovskii functional \cite{brazovskii:75:0} in the Gaussian approximation
  \begin{equation}
  \label{LB}
    {\cal L}_G[\varPhi]\approx\frac{c}{2}\int_V \frac{d{\bf k}}{(2\pi)^d} 
    \tilde \varPhi({\bf k}) [t_0+(k^2-k_0^2)^2]
\tilde \varPhi(-{\bf k})
   \end{equation}
 where $k_0$ is given in (\ref{k0}),
 \begin{equation}
 \label{t0}
  t_0=\tilde C(k_0)/c,
 \end{equation}
 $\tilde C(k_0)$ is given in (\ref{Ck0}) and $c=\beta^*\lambda_D^{*2}/\sqrt{a_Na_D}$.  
  The Landau-Brazovskii functional was successfully used for a description of the structure of 
  block copolymers \cite{leibler:80:0}, binary or ternary surfactant mixtures \cite{gompper:94:0,gozdz:96:1} 
  and recently for the colloidal self-assembly \cite{ciach:13:03}. In each case the physical interpretation of 
  the order-parameter $\varPhi$ is different.
  In particular, in the case of ternary surfactant mixtures $\varPhi$ is interpreted as a concentration 
  difference between the polar and nonpolar components, in close analogy with the   present case where $\varPhi$ is 
  a concentration 
  difference between the inorganic and organic components.
  
  From Eq.(\ref{C}) we can easily obtain 
  the approximate expression for the correlation function in the real-space representation \cite{ciach:01:2},
  \begin{equation}
\label{Cr}
  G(r)=\frac{\lambda}{2\pi r}e^{-r/\xi}\sin\big(\frac{2\pi r}{\lambda}\big),
 \end{equation}
 where the dimensionless correlation length and  period of the exponentially 
 damped oscillatory decay, $\xi$ and $\lambda$, are given by
  \begin{equation}
   \xi^2=\frac{2}{\sqrt{t_0+k_0^4}-k_0^2}
  \end{equation}
  and
\begin{equation}
   \Big(
   \frac{\lambda}{2\pi}
   \Big)^2=\frac{2}{\sqrt{t_0+k_0^4}+k_0^2}.
  \end{equation}
 Note that $\xi$ differs from $\xi_0$ and diverges  at the boundary of stability of the disordered phase, 
 $t_0=0$. Moreover,  $\lambda\le 2\pi/k_0$; the equality holds 
 when $t_0=0$.
 
 We should mention that in the  Brazovskii theory
 the dominant fluctuations ($k\simeq k_0$) lead to
 fluctuation-induced first-order phase transition to the lamellar phase beyond the Gaussian approximation 
 ~\cite{brazovskii:75:0}, or as shown recently to
 a transition to a nematic phase \cite{barci:13:0}. The $k$-dependence of $\tilde G(k)$ in the disordered phase 
 remains almost unchanged beyond the Gaussian approximation, however, but 
 $t_0$, and hence the model parameters,  become renormalized
~\cite{brazovskii:75:0}. For this reason  Eq.(\ref{G}) can 
 reasonably well describe the
 experimental results, but the MF theories cannot correctly predict the 
 microscopic expressions for the parameters present in Eq.(\ref{G}). 
 
 The similarity between our system and ternary surfactant mixtures was confirmed by the agreement of the experimental
 results for the former and the formulas developed for the latter case \cite{sadakane:13:0}. 
 Here we have shown that starting from a simple microscopic density functional theory we can obtain by 
 a systematic coarse-graining  procedure
 the same Landau-Brazovskii functional (\ref{LB})
  that  describes the amphiphilic systems. 

\section{Effects of confinement}
 \subsection{The Euler-Lagrange equations}
 
 In  order to obtain
 the form of $\Phi$ and $\rho_q^*$ it is necessary to solve the EL equations for $\Phi$, $\psi$, $\rho_{ion}^*$ 
 and $\rho_q^*$. 
 The expressions corresponding to the full functional  (\ref{OmegaL}) 
 with the electrostatic and the vdW
 contributions to the internal energy given in Eqs.(\ref{el}) and (\ref{uvdW}) respectively,
 and the entropy given in (\ref{S})
 are rather long, and are given in Appendix A. 
 Here we discuss the linearized EL equations for $\bar\Phi=0$ (see (\ref{EL1})
 and (\ref{EL2}) in the Appendix A) that can be written in the form
\begin{equation}
\label{Philin}
\nabla^2\nabla^2\Phi({\bf r})+a_2\nabla^2\Phi({\bf r})+a_0\Phi({\bf r})=0
\end{equation}
and
 \begin{equation}
\label{EL1l}
\rho_q^*(z)=\frac{(1-\bar \rho_{ion}^*)}{T^*}\Bigg[\xi_0^{-2} \Phi(z)-\nabla^2\Phi(z)\Bigg] ,
\end{equation}
where $\nabla^2$ denotes the laplacian,
\begin{equation}
\label{a2}
 a_2
 =[(a_N-a_D)-\lambda_D^{*2}\xi_0^{-2}]\lambda_D^{*-2},
\end{equation} 
and
\begin{equation}
 a_0
 =a_D\xi_0^{-2}\lambda_D^{*-2}.
\end{equation}
 
Let us focus for simplicity on one-dimensional concentration profiles. 
The solution of (\ref{Philin}) for $\varPhi$ depending only on $z$ is a linear combination of 
the exponential terms $\exp(\pm\lambda_iz)$ with
\begin{equation}
\label{delta12}
 \lambda_{1,2}^2=\frac{-a_2\pm\sqrt\Delta}{2}
\end{equation}
where $\Delta$ can be written in the form 
\begin{equation}
\label{Delta}
 \Delta =\Big[
 \lambda_D^{*2}\xi_0^{-2}-(\sqrt a_N-\sqrt a_D)^2
 \Big]\Big[
 \lambda_D^{*2}\xi_0^{-2}-(\sqrt a_N+\sqrt a_D)^2
 \Big]\lambda_D^{*-4}.
\end{equation}
Note that $\Delta=0$ when either the first or the second factor in (\ref{Delta}) vanishes. 
The first factor in (\ref{Delta}) is proportional to $t_0$ (see (\ref{t0})), 
hence $\Delta$ vanishes at the $\lambda$-surface $t_0=0$. 
The second factor vanishes at the disorder surface 
$\lambda_D^{*}/\xi_0=(\sqrt a_N+ \sqrt a_D) $.
For $(\sqrt a_N-\sqrt a_D)^2<(\lambda_D^{*}/\xi_0)^2<(\sqrt a_N+\sqrt a_D)^2$  
the  $\lambda_i$  are complex conjugate numbers, because  $\Delta<0$. 
For $\lambda_D^{*}/\xi_0>(\sqrt a_N+ \sqrt a_D) $ 
 both $\lambda_i$ are real, while
for $(\lambda_D^{*}/\xi_0)^2<(\sqrt a_N-\sqrt a_D)^2$
 (i.e. $t_0<0$)
both $\lambda_i$ are imaginary.  Note 
that for $t_0=0$ we obtain an oscillatory function 
with the wavenumber of oscillations equal to $k_0$, in consistency with the results of the previous section.


\subsection{The boundary conditions for the EL equations}
When the system is in contact with a selective and charged surface at $z=0$, then there is additional
contribution to the internal energy~\cite{ciach:10:0,pousaneh:12:0,pousaneh:14:0}
\begin{equation}
  U_s=\Big[J\Big(\frac{\Phi^2(0)}{2}-h_0\Phi(0)
 \Big) + e\sigma_0\psi(0)\Big] A^*,
\end{equation}
where $A^*$ is the dimensionless area of the confining surface, $\sigma_0$ is the dimensionless  surface charge
(the charge per area $a^2$), 
$h_0$ is the dimensionless surface field describing the preferential adsorption
of the inorganic (for $h_0>0$) or organic (for $h_0<0$) components, and the first term follows from the missing
fluid neighbors for $z\le 0$. In the case of a slit with another surface at $z=L$  there is analogous contribution
to the internal energy. The above surface terms lead to the boundary conditions for the EL equations 
\cite{ciach:10:0,pousaneh:14:0}
\begin{equation}
\label{BCel}
 \nabla\Phi(0)-\Phi(0)=-h_0, \hskip2cm  \nabla\Phi(L)+\Phi(L)=h_L
\end{equation}
and
\begin{equation}
\label{BCel1}
 \nabla \psi (0)=-\frac{4\pi e}{\epsilon}\sigma_0,\hskip2cm    \nabla \psi (L)=\frac{4\pi e}{\epsilon}\sigma_L.
\end{equation}
In the linearized theory 
\begin{equation}
 \psi({\bf r})=\frac{k_BT}{e(1-\bar\rho_{ion}^*)}\Big[\Phi({\bf r})-\frac{\rho_q^*({\bf r})}{\bar\rho_{ion}^*}\Big]
\end{equation}
 when $\bar\Phi=0$, and from  (\ref{EL1l}) 
we obtain using (\ref{BCel1})
the second boundary conditions for the equation (\ref{Philin})
\begin{equation}
\label{BC2}
 \Phi^{'''}(0)-\zeta^{-2}\Phi'(0)
 =-\frac{T^*}{\lambda_D^{*2}}\sigma_0,\\ \hskip2cm  
 \Phi^{'''}(L)-\zeta^{-2}\Phi'(L)
 =\frac{T^*}{\lambda_D^{*2}}\sigma_L
\end{equation}
where 
\begin{equation}
\label{zeta}
 \zeta^{-2}=\xi_0^{-2}-\frac{T^*\bar\rho_{ion}^*}{1-\bar\rho_{ion}^*}
\end{equation}

\subsection{The local concentration and the local charge density in a semiinfinite system}

We shall limit ourselves to a near-surface structure in the disordered phase and 
to the critical composition $\bar\Phi=0$. 
The composition of the near-surface layer depends on both the wall-fluid van der Waals interactions $h_0$ and
on the sign and value of the surface charge $\sigma_0$. A hydrophilic surface attracts water, and
 a negatively  charged surface 
attracts water-soluble ions. In contrast, the positively charged surface attracts ions soluble in the organic solvent,  
and in the case of the positively charged hydrophilic surface the composition in its vicinity
depends on the ratio $h_0/\sigma_0$. We shall limit ourselves to $h_0>0$.

Let us first focus on the case of
$\lambda_D^{*}/\xi_0>(\sqrt a_N+ \sqrt a_D) $, where
 both $\lambda_i$ are real numbers. From (\ref{Philin}) and (\ref{EL1l})  we have
 \begin{equation}
 \label{Phi(z)}
  \Phi(z)=A_1e^{-\lambda_1 z}+A_2e^{-\lambda_2 z}
 \end{equation}
 and
 \begin{equation}
 \label{rhoq(z)}
  \rho_q^*(z)=\frac{(1-\bar \rho_{ion}^*)}{T^*}\Bigg[
  (\xi_0^{-2}-\lambda_1^2)A_1e^{-\lambda_1 z}+(\xi_0^{-2}-\lambda_2^2)A_2e^{-\lambda_2 z}\Bigg]
 \end{equation}
 where $A_i$ are determined by the boundary conditions and depend on $\sigma_0$ and $h_0$ (see Appendix C).
 Note that the asymptotic decay at large separation of both the excess concentration and the charge density is given by
 the inverse decay length $\min(\lambda_1,\lambda_2)$, while in the case of the hydrophilic salt the decay length of 
 the excess concentration is $\xi_0$, and the decay length of the charge is $\lambda_D^*$. This difference follows from
 the coupling of the concentration and charge fluctuations already in the Gaussian approximation in the case of the 
 antagonistic salt.

  The two decay lengths approach $\xi_0$ and $\lambda_D^*$ only far away from the disorder line, i.e. 
 for 
 $\lambda_D^{*}/\xi_0\gg(\sqrt a_N+ \sqrt a_D) $ (low ionic strength, away from the critical point). 
 In the limit of $\xi_0/\lambda_D^{*}\to 0$ Eqs.(\ref{Phi(z)}) and (\ref{rhoq(z)}) take the forms
 \begin{equation}
  \Phi(z)\simeq h_0e^{-z/\xi_0}
 \end{equation}
and
 \begin{equation}
  \rho_q^*(z)\simeq -\frac{\sigma_0}{\lambda_D^*}e^{-z/\lambda_D^*},
 \end{equation}
where we took into account that $a_D\approx 1$.

 The difference between the two decay lengths decreases when the disorder line  
 ($\lambda_D^{*}/\xi_0=(\sqrt a_N+ \sqrt a_D) $) is approached, and
at the disorder line they become identical,
 \begin{equation}
  \lambda_1^2=\lambda_2^2
  =\frac{\sqrt a_D}{\lambda_D^*\xi_0}.
 \end{equation}
 
 
 For $(\sqrt a_N- \sqrt a_D)<\lambda_D^{*}/\xi_0<(\sqrt a_N+ \sqrt a_D) $ 
 Eq.(\ref{Phi(z)}) can be written inthe form
 \begin{equation}
  \Phi(z)=A\cos(\lambda_{im}z+\theta)e^{-\lambda_{re}z}
 \end{equation}
with similar damped oscillatory decay of $\rho_q^*(z)$, because 
 $\lambda_1=\lambda_{re}+i\lambda_{im}$ and $\lambda_2=\lambda_{re}-i\lambda_{im}$ are                                
 complex conjugate numbers.

 \subsection{The structure and effective potential between parallel surfaces}
\begin{figure}[h]
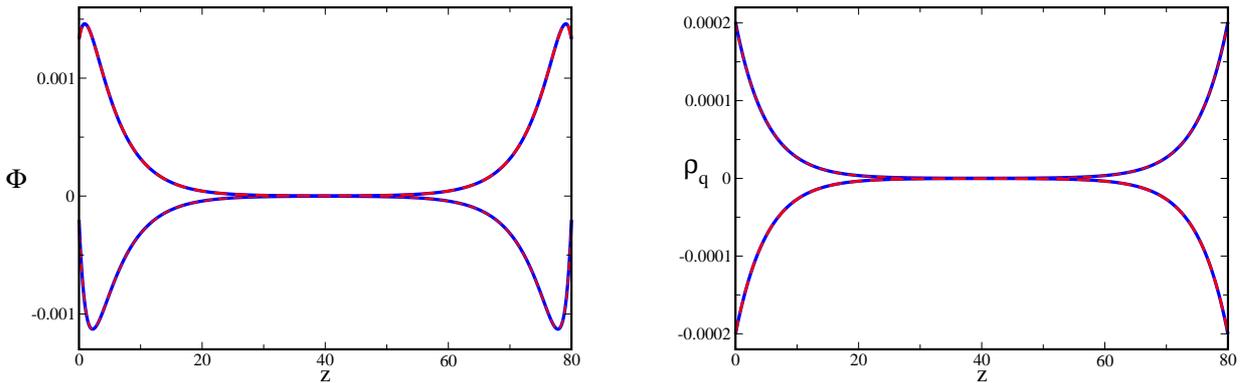

\center
\vspace{2cm}
 \includegraphics[scale=0.3]{./Figures/Fig1_a.eps}
 \hspace{1cm}
  \includegraphics[scale=0.3]{./Figures/Fig1_b.eps}
\caption{ The dimensionless 
concentration profile defined in Eq.~\ref{OP} (left) and the dimensionless 
charge density (right). Solid lines correspond to numerical results of the full
EL equations Eqs.~(\ref{rhoelim})-(\ref{EL3}) and (\ref{Poisson}), while dash lines show analytical results  
(Eqs.~(\ref{Phi(z)}), (\ref{rhoq(z)})). Dimensionless parameters are $\lambda_D^*= 5$, $|T/T_c-1|= 0.1$, $\bar \rho_{ion}^* = 0.001$,  $h= 0.001$,
 $\sigma= -0.001$ for top curves and $\sigma= 0.001$  for bottom curves. The distance from the left wall $z$ is in 
units of the microscopic length $a\approx 0.4 nm$.
}
\label{OP1}
\end{figure}

\begin{figure}[h]
\center
\vspace{2cm}
 \includegraphics[scale=0.3]{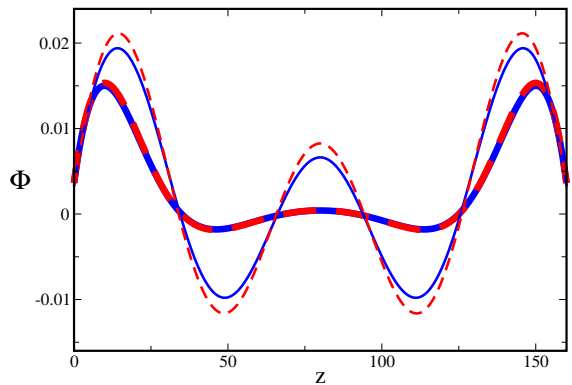}
\caption{ The dimensionless 
concentration profile. Solid lines correspond to numerical results of the full EL 
equations Eqs.~(\ref{rhoelim})-(\ref{EL3}) and (\ref{Poisson}), while dash lines show analytical 
results,  Eq.~\ref{Phi(z)}. The dimensionless parameters are $h= 0.001$, $\sigma= -0.001$,  $\lambda_D^*= 10$, 
$\bar \rho_{ion}^* = 0.005$, (thin curves) $ |T/T_c-1|= 0.002$, and (thick curves)  $|T/T_c-1|= 0.005$.
The distance from the left wall $z$ is in 
units of the microscopic length $a\approx 0.4 nm$.
}
\label{OP2}
\end{figure}

\begin{figure}[h]
\center
\vspace{2cm}
 \includegraphics[scale=0.3]{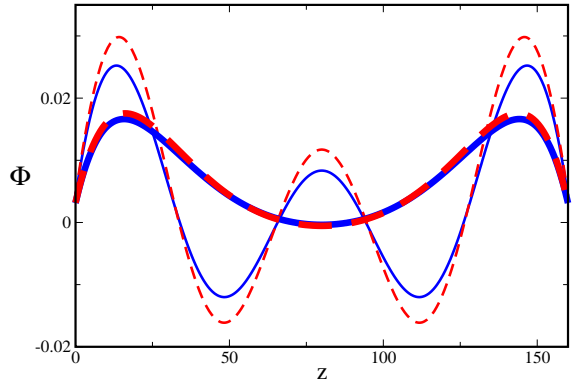}
\caption{The dimensionless 
concentration profile. Solid lines correspond to numerical results of the full EL 
equations Eqs.~(\ref{rhoelim})-(\ref{EL3}) and (\ref{Poisson}), while dash lines show analytical 
results,  Eq.~\ref{Phi(z)}.  The dimensionless parameters are $h= 0.001$, $\sigma= -0.001$,  
$ |T/T_c-1|= 0.001$, (thick curves) 
$\lambda_D^*= 20$,  $\bar \rho_{ion}^* = 0.00125$, and (thin curves) $\lambda_D^*= 10$, $\bar \rho_{ion}^* = 0.005$. 
The distance from the left wall $z$ is in 
units of the microscopic length $a\approx 0.4 nm$.
}
\label{OP3}
\end{figure}

\begin{figure}[h]
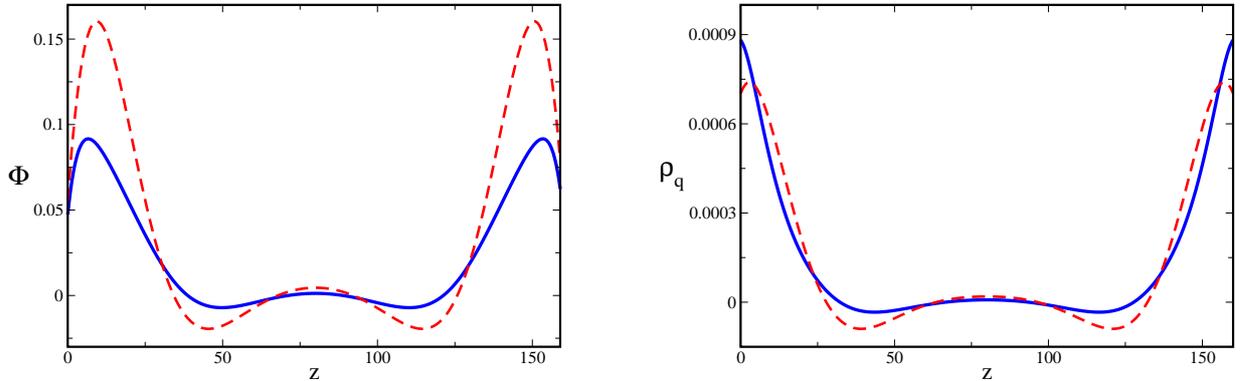

\center
\vspace{2cm}
 \includegraphics[scale=0.3]{./Figures/Fig4_a.eps}
 \hspace{1cm}
  \includegraphics[scale=0.3]{./Figures/Fig4_b.eps}
\caption{ The dimensionless 
concentration profile (left) and the dimensionless 
charge density (right). The dimensionless  parameters are $h= 0.03$, $\sigma= -0.01$, 
$\lambda_D^*= 10$, $\bar \rho_{ion}^* = 0.005$ and  $| T/T_c-1|= 0.002$. 
The distance from the left wall $z$ is in 
units of the microscopic length $a\approx 0.4 nm$.}
\label{OP4}
\end{figure}

 We consider two selective and charged surfaces ($h_{0}, \sigma_0$ and $h_L,\sigma_L$)
 which are separated by the distance L. We limit ourselves to identical surfaces, with $h=h_0=h_L$ and 
 $\sigma=\sigma_0=\sigma_L$.
 The local concentration has the form
 \begin{equation}
 \label{OPcon}
  \Phi(z)={\cal A}_1[e^{-\lambda_1 z}+e^{-\lambda_1 (L-z)}]+{\cal A}_2[e^{-\lambda_2 z}+e^{-\lambda_2 (L-z)}]
 \end{equation}
 in the case of the structureless fluid, and 
 \begin{equation}
 \label{OPcon1}
  \Phi(z)={\cal A}[\cos(\lambda_{im}z+\vartheta)e^{-\lambda_{re}z}+\cos(\lambda_{im}(L-z)+\vartheta)e^{-\lambda_{re}(L-z)}]
 \end{equation}
 in the presence of mesoscopic inhomogeneities. The expressions for the amplitudes and the phase are 
 rather long and will not be given here.
 The results of the linear theory for  the excess concentration profiles, Eqs.(\ref{OPcon}) and (\ref{OPcon1}),
are compared with the numerical solutions of the full EL equations, Eqs.~(\ref{rhoelim})-(\ref{EL3}) 
and (\ref{Poisson}) for $T^*=6$ in Figs.~\ref{OP1}-\ref{OP4}.

As Fig.~\ref{OP1} shows, for small surface charges, surface fields and ion concentrations, and 
for temperatures far from the critical point, the linear theory agrees very well with the numerical solutions
of the EL equations. 
Figs.~\ref{OP2} and \ref{OP3} show the appearance of the periodic structure upon approaching 
the critical point and upon increasing $\lambda_D^*$.
Finally, Fig.~\ref{OP4} indicates that for
bigger surface charges and surface fields the linear theory differs significantly from the numerical results,
however the qualitative agreement is preserved.\\

 The  excess grand potential (\ref{L0}) 
 consists of the surface tension contribution, $(\gamma_0+\gamma_L)A^*$, and the effective 
 potential between the confining surfaces of the dimensionless area $A^*$,  $A^*\Psi(L)$. 
 When the concentration $\Phi$ and the charge density
  $\rho_q^*$ satisfy 
 the EL equations (\ref{EL2}) and (\ref{EL1}), then
 \begin{eqnarray}
 \label{pot}
  \gamma_0+\gamma_L+\Psi(L)=-\frac{J}{2}\Bigg[\Big(h_0 -\frac{\sigma_0T^*}{1-\bar\rho_{ion}^*}\Big)\Phi(0)+
 \Big( h_L-\frac{\sigma_LT^*}{1-\bar\rho_{ion}^*}\Big)\Phi(L)\\
 \nonumber
  +\frac{T^*}{\bar\rho_{ion}^*(1-\bar\rho_{ion}^*)}\Big(\sigma_0\rho_q(0)+\sigma_L\rho_q(L)\Big)\Bigg].
 \end{eqnarray}
The results of the previous section give us for $\lambda_D^{*}/\xi_0>(\sqrt a_N+ \sqrt a_D) $ 
 \begin{equation}
 \label{Psi(L)}
  \Psi(L)=C_1e^{-\lambda_1 L}+C_2e^{-\lambda_2 L}
 \end{equation}
 and for $(\sqrt a_N- \sqrt a_D)<\lambda_D^{*}/\xi_0<(\sqrt a_N+ \sqrt a_D) $ the above can be written in the form
 \begin{equation}
  \Psi(L)=C\cos(\lambda_{im}L+\theta)e^{-\lambda_{re}L}
 \end{equation}
  where the expressions for the amplitudes are too long to be given here. 
 
\begin{figure}
\center
\vspace{2cm}
 \includegraphics[scale=0.3]{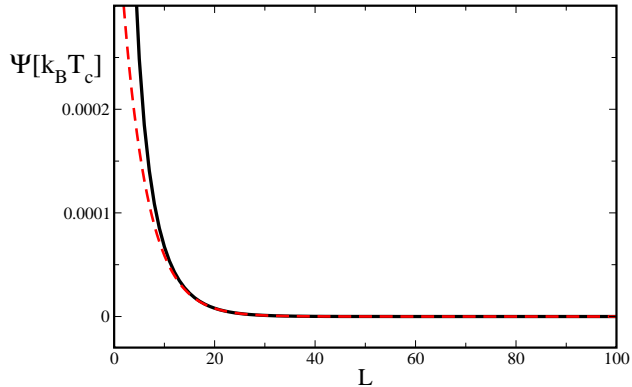}
\caption{The effective potential per microscopic area $a^2$ between two identical surfaces.
Solid line corresponds to numerical
solutions of the full EL equations Eqs.~(\ref{rhoelim})-(\ref{EL3}) and (\ref{Poisson}) and dash line corresponds
to  Eq.(\ref{pot}).  The dimensionless parameters are $h= 0.001$, $\sigma= 0.001$,  $\lambda_D^*= 5$, 
$ |T/T_c-1|= 0.1$ and $\bar \rho_{ion}^* = 0.001$. 
}
\label{pot.1}
\end{figure}

\begin{figure}[h]
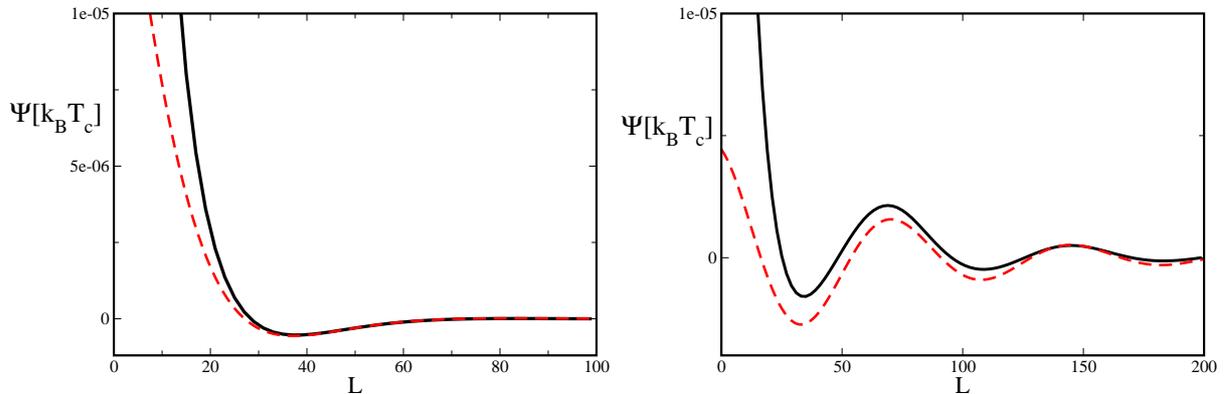

\center
\vspace{2cm}
 \includegraphics[scale=0.3]{Figures/Fig6_a.eps}
  \includegraphics[scale=0.3]{Figures/Fig6_b.eps}
\caption{ The effective potential per microscopic area $a^2$ between two identical surfaces.
Solid line corresponds to numerical
solutions of the full EL equations Eqs.~(\ref{rhoelim})-(\ref{EL3}) and (\ref{Poisson}) and dash line corresponds
to  Eq.(\ref{pot}). The dimensionless  parameters are $h= 0.001$, $\sigma= -0.001$,  
$\lambda_D^*= 15$, $\bar \rho_{ion}^* = 0.005$ and (Left) $|T/T_c-1|= 0.005$ and (right) $ |T/T_c-1|= 0.002$.}
\label{pot.2}
\end{figure}

  In Figs.~\ref{pot.1}-\ref{pot.2} the effective potential per microscopic area $a^2$
   between identical surfaces, obtained in the linearized and the nonlinear theory is presented for different 
  $\lambda_D^*$, $T^*$ and different surface charge and selectivity of the surfaces. 
    Fig.~\ref{pot.1} presents a very good agreement at large distances between the
linear theory and the full EL equations for small surface fields and surface charge, and for temperatures  far 
from the critical point. When the critical 
point of the binary mixture is approached and  the disorder surface is crossed, 
an oscillatory force between the surfaces is seen 
(Fig.~\ref{pot.2} (right)).
Note that for all the considered cases there is  qualitative agreement between the linear theory 
and the numerical results.

The potential between surfaces of  area $400nm\times400nm \approx 10^6a^2 $ is $10^6\Psi(L)$. 
Note that for such a mesoscopic 
surface the first two extrema in Fig.\ref{pot.2} are both of order of $k_BT$.

 \section{the case of  experimentally studied samples}
 
   \label{cexp}
In this section we apply our theory to the systems studied experimentally in 
Ref.\cite{sadakane:13:0,sadakane:09:0,leys:13:0}.
In the experiments the antagonistic salt sodium tetraphenylborate (NaBPh$_4$) 
was added to the $3-$methylpyridine (3MP) and heavy water
mixture near its lower critical point (LCP).  Small-angle neutron scattering (SANS) 
was performed and a periodic structure with a periodicity of about $10 nm$ 
 was reported in the experiment.  The structure factor of the ternary mixture was 
 determined for a few salt concentrations and a few
temperatures. The structure factor was fitted to the expression obtained by 
Onuki and Kitamura \cite{onuki:04:0} that has the same $k$ 
dependence as our Eq.(\ref{G}). 

\begin{figure}
\center
 \includegraphics[scale=0.3]{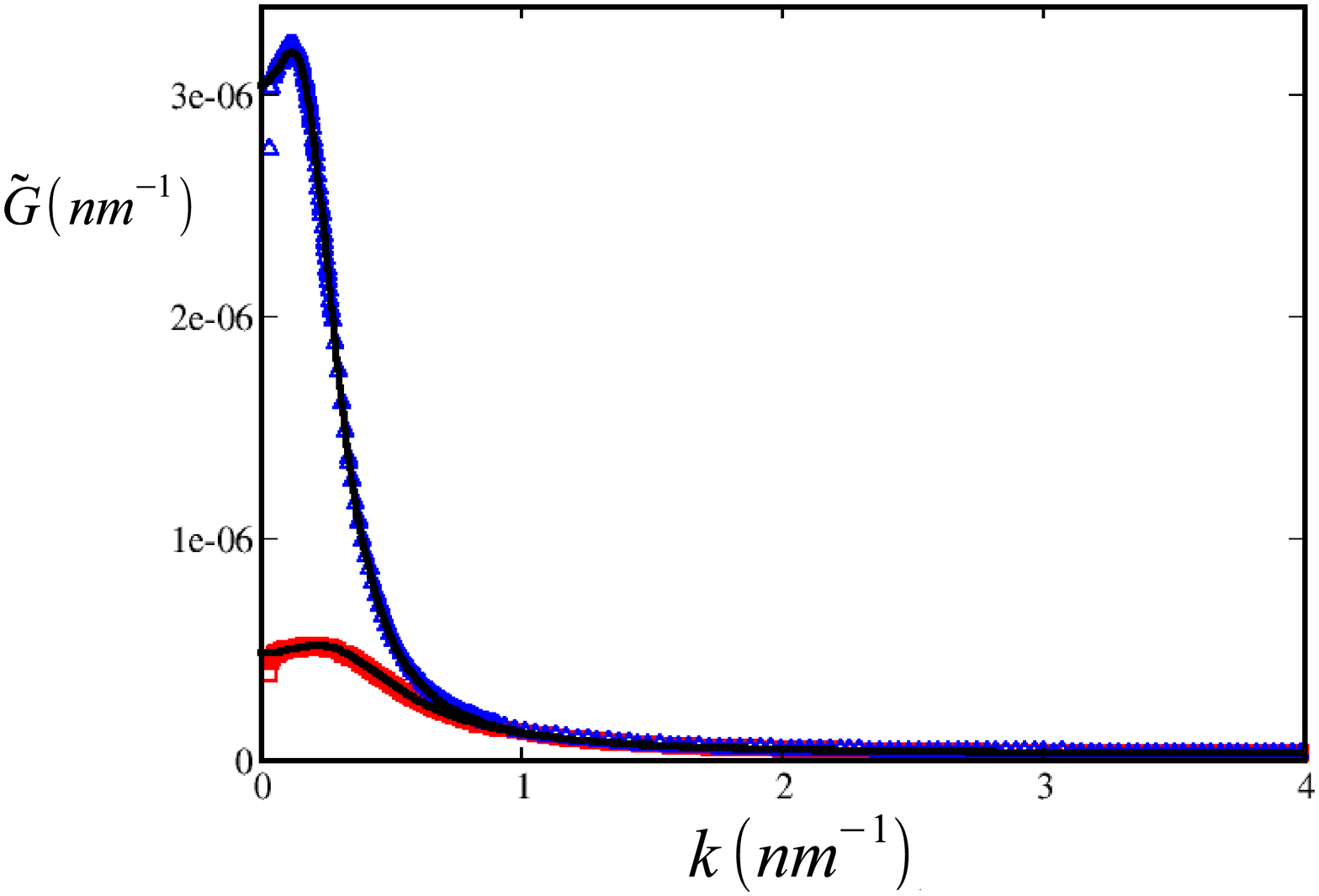}
  \includegraphics[scale=0.3]{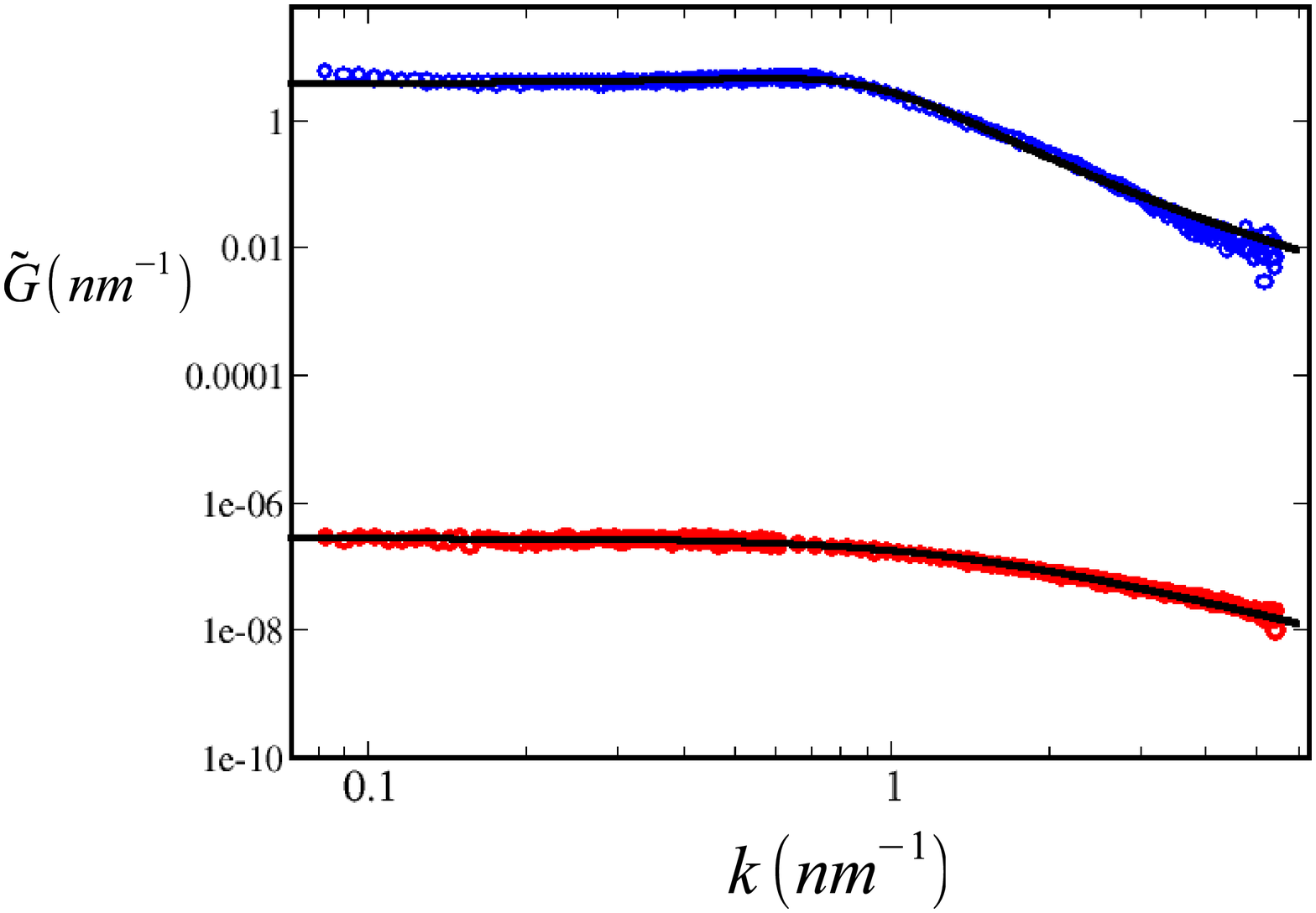}
\caption{  SANS intensity measurements for near-critical composition of the 3MP, D$_2$O and NaBPh$_4$ mixture. 
Top:
$T= 293K$ (bottom curve), 
$T= 313 K$ (top curve) and $\rho_{ion}=7.2mM/L$ \cite{leys:13:0}. Bottom:  $T=280K$,
$\rho_{ion}= 300mM/L$ (top curve),  and $\bar\rho_{ion}= 1mM/L$ (bottom curve)\cite{sadakane:13:0}. The 
remaining parameters are shown in Table~\ref{tab1}.
The solid lines are the theoretical prediction of the correlation function, Eq.(\ref{G}). 
The fit for $\bar\rho_{ion}= 300mM/L$ is  better than in Ref.\cite{sadakane:13:0}
because  we take into account the dependence of $a_D$ on $\bar\rho_{ion}$.}
\label{fitLey}
\end{figure}

Our theory has been developed for the upper critical point. In order to apply it to
the system that phase separates for increasing temperature,
we assume that in the coarse-grained description of the considered mixture
the interaction parameter $J$ depends on $T$ in such a way that near the LCP
 $T^*=k_BT/J(T)$ decreases for increasing $T$. 
 We shall not attempt to reproduce the phase diagram. Our purpose is a calculation of the effective potential 
 between the confining surfaces for the samples studied  experimentally in 
 Refs.\cite{sadakane:13:0,sadakane:09:0,leys:13:0}.
 In order to model the particular samples, we  assume that $a=0.4 nm$, $\bar\Phi=0$ for the samples with near-critical
 composition,
 $\lambda_D^*$ is given in Eq.(\ref{lambdaD}) and we 
 fit Eqs.(\ref{G})-(\ref{aD}) to the measured scattering intensity $S(k)$, assuming that $S(k)=S_0\tilde G(k)$.
 The remaining parameters are taken from Ref.\cite{sadakane:13:0,leys:13:0}.
 The selected samples and the   parameters obtained from the fitting are given in  Table I, 
 and the fitting of the formulas (\ref{G})-(\ref{aD}) to the experimental curves is shown in Fig.\ref{fitLey}.
  
 \vspace{1cm}
\begin{table}[h]
\hspace{-0.1cm}
\begin{tabular}{ l@{\hskip 0.2in} l@{\hskip 0.2in}  l@{\hskip 0.2in}  l@{\hskip 0.2in}   c @{\hskip 0.2in} c@{\hskip 0.2in} l@{\hskip 0.2in}  c l@{\hskip 0.2in}}
\hline
\vspace{0.1cm}
 $T (K) $&$\rho_{ion}(mM/L)$&$\lambda_D(nm)$&$\xi_0 (nm)$&$ a_N$&$S_0\tilde G_{\Phi\Phi}(0)$ &$T^*$ & $\Lambda_1  (nm)$ & $\Lambda_2  (nm)$ \\
\hline
313&7.2  &2.6&5.82&$ 1.20$&$1.42\cdot 10^{-8}$ &$106$ &$5.5$& $ 34.6$\\
293&7.2  &2.6&2.16& $ 1.73 $&$1.55\cdot 10^{-8}$ &$152$ &$2.94$ & $ 25.1$\\
\hline
280&300 &0.3&$3.89$&$ 1.05$&$0.03875$ & 160&1.90& 8.26\\
280&1 &5.2&$0.75$& $ 1.03 $&$7.59\cdot 10^{-8}$ &160&0.76&5.16\\
\hline
\label{fitLeytab}
\end{tabular}
\caption{
Parameters characterizing the samples studied in Ref.\cite{leys:13:0} (the first two rows) and in  Ref.\cite{sadakane:13:0}
 (the last two rows). The data in the first three columns are taken or computed from the data given in
Refs.\cite{leys:13:0,sadakane:13:0}. The data in the columns 4-6 are obtained from the fitting of the SANS intensity to
 the correlation function $\tilde G$ given in Eq.(\ref{G})-(\ref{aD}).
 The data in column 7 are computed from Eq.(\ref{aN}).  For the first three rows
 the columns 8 and 9 show the decay length and the period
 of the damped oscillatory decay, i.e.
 $\Lambda_1=1/\lambda_{re}$ and $\Lambda_2=2\pi/\lambda_{im}$ (Eq.\ref{lambdare}).
 In the last row (1 mM/L sample) the two decay lengths
 of the double exponential decay are shown, i.e. $\Lambda_i=1/\lambda_i$ (see Eq.(\ref{delta12})).
 }
\label{tab1}
\vspace{1cm}
\end{table}

\begin{figure}
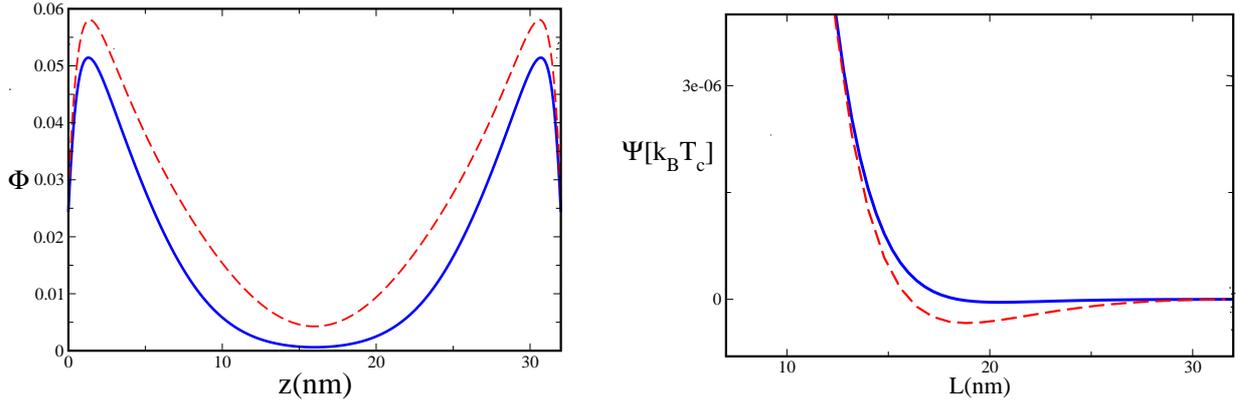

 \includegraphics[scale=0.3]{Figures/Fig8_a.eps}
  \hspace{0.5cm}
    \includegraphics[scale=0.3]{Figures/Fig8_b.eps}
  \caption{
  (Left) Excess concentration profile (dimensionless) and (right) the effective potential 
   per microscopic area $a^2$ (with $a\approx 0.4nm$) resulted
  from full EL equations, (\ref{rhoelim})-(\ref{EL3}) and (\ref{Poisson}), as a function of 
  the distance between two identical surfaces.
   The parameters are selected according to experiment \cite{leys:13:0} for  $7.12  mM/L$  of  $NaBPh_4$ added to $D_2O$
 and metylpyridine at $313 K$ (dash line) and $293 K$ (solid line) given in Table.~\ref{tab1}. 
 The dimensionless selectivity and surface charge are $h=0.001$ and  $\sigma=0.001$ respectively.
  }
    \label{OPLey}
 \end{figure}

\begin{figure}
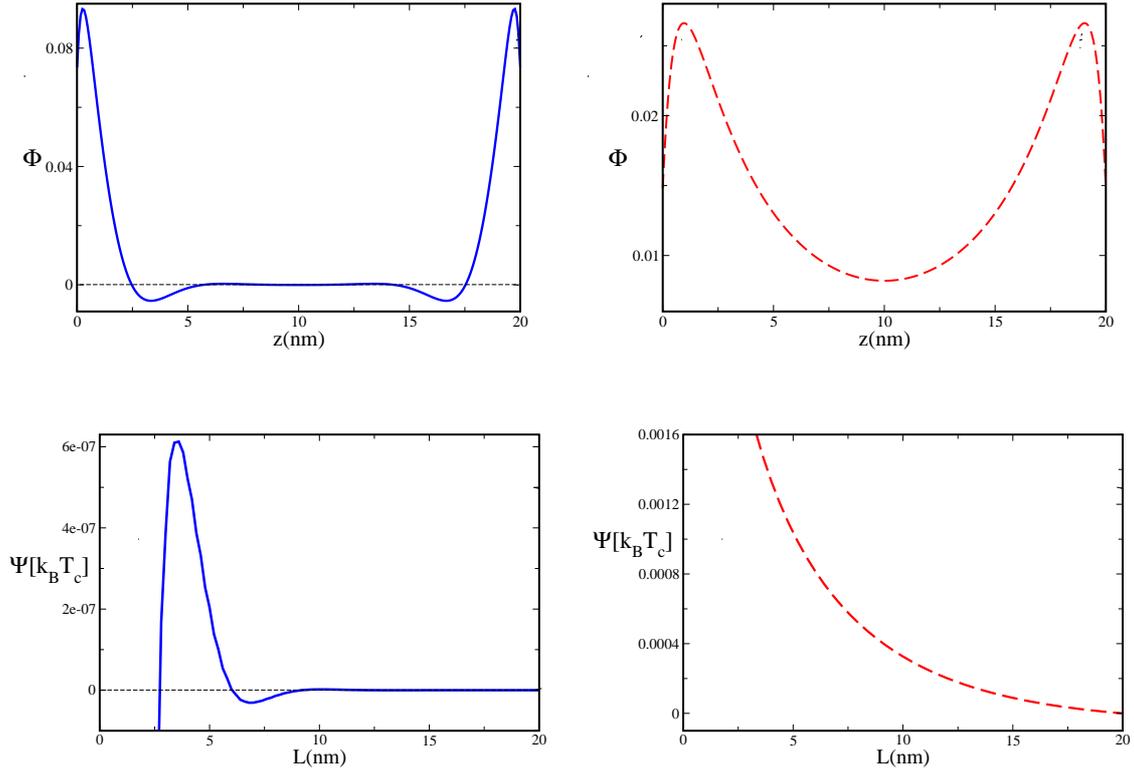

   \hspace{-0.5cm}
  \includegraphics[scale=0.27]{./Figures/Fig9_a.eps}
    \hspace{0.5cm}
  \vspace{1.cm}
    \includegraphics[scale=0.27]{./Figures/Fig9_b.eps}
       \hspace{0cm}
     \includegraphics[scale=0.26]{./Figures/Fig9_c.eps}
          \hspace{0.2cm}
  \hspace{0.cm}
    \includegraphics[scale=0.26]{./Figures/Fig9_d.eps}
     \hspace{0.2cm}
  \caption{ (Top row) Excess concentration profiles (dimensionless) and (bottom row) the effective potential
  per microscopic area $a^2$ (with $a\approx 0.4nm$)
  resulted from full EL equations, (\ref{rhoelim})-(\ref{EL3}) and (\ref{Poisson}), as a function of the
  distance between two identical surfaces. 
   The parameters are selected according to experiment \cite{sadakane:13:0}  for the $300 mM/L$ 
   (solid lines) and  $1 mM/L$ (dash lines)
  of  $NaBPh_4$ added to $D_2O$
 and metylpyridine
  sample at $T=280K$~\cite{sadakane:13:0}, given in Table.~\ref{tab1}. 
  The dimensionless selectivity and surface charge are $h=0.001$ and  $\sigma=0.001$ respectively.}
     \label{OPSadakane}
 \end{figure}

 The excess  concentration profiles and the effective potential for the
 samples studied in Ref.\cite{leys:13:0} and in  Ref.\cite{sadakane:13:0} are shown in Figs.~\ref{OPLey}
 and \ref{OPSadakane} respectively.
 We plot the results of the full EL equations, (\ref{rhoelim})-(\ref{EL3}) and (\ref{Poisson}),
  between two identical weakly selective and weakly charged surfaces,  for the
  parameters  given in Table.~\ref{tab1}. 
  To obtain $ |T/T_c-1|$ from $\xi_0$
 we use the  critical exponent $\nu=0.63$ instead of $1/2$ in Eq.(\ref{xi}).
The structure factor takes a maximum for $k=k_0>0$ for three of the selected samples, 
and for $k_0=0$ for the low salt concentration, $\rho_{ion}=1 mmol/L$.  For $k_0>0$ the exponentially damped oscillatory
decay of correlations occurs, but the decay length is short compared to the period of oscillations (table I).
For this reason the second extremum is much smaller than the first one, and is not seen on the plot.
In the $k_0=0$  case the 
 decay of correlations is given by two exponential functions.

 \vspace{1cm}
 
 Note the change of the shape of $\Psi(L)$
   with increased amount of salt (Fig.\ref{OPSadakane}). Because of the very short screening length the electrostatic 
   repulsion that dominates in the $1mM/L$ sample is suppressed.
   In the $300mM/L$ sample  a repulsion barrier $\sim 0.6 k_BT$ for surfaces of area 
   $1.6\cdot 10^5nm^2$ at the separation
   $L\approx 5 nm$ occurs, and the barrier is followed by a strong attraction for $L<3nm$. For
   the intermediate salt concentration, $\bar\rho_{ion}=7.2 mM/L$ 
   a minimum of the potential of the depth  $\sim 0.5 k_BT$ for surfaces of area 
   $1.6\cdot 10^5nm^2$ occurs for $L\simeq 20nm$.
   
 \section{Discussion and summary}
 
 We have developed a Landau-type functional for a near-critical mixture with addition of antagonistic salt.
 We used the same coarse-graining procedure as in the case of the inorganic
 salt~\cite{ciach:10:0,pousaneh:12:0,pousaneh:14:0}. In both cases we have postulated that the OP 
 for the phase separation, $\Phi$, is the concentration difference between the inorganic and organic components.
 While in the case of the inorganic salt $\Phi=\rho_w^*-\rho_l^*+\rho_{ion}^*$, in the case of the antagonistic salt 
 $\Phi=\rho_w^*-\rho_l^*+\rho_q^*$. Since there are two types of ions, we have postulated that the entropy has a form of the
 ideal entropy of mixing of a four-component mixture.  The entropy of mixing has quite different form for the
 two different OP when expressed in terms of $\Phi$, $\rho_{ion}^*$ and $\rho_q^*$.
 As a  consequence,   the linearized EL equations 
  for $\Phi$ and $\rho_q^*$ are coupled in the case of the antagonistic salt, and decoupled in the case of the 
  inorganic salt.
  In the absence of the coupling of the  linearized EL equations the decay lengths of $\Phi$ and $\rho_q^*$ are
  $\xi_0$ and $\lambda_D^*$ respectively.
  In the presence of the coupling the decay of both $\Phi(z)$ and $\rho_q^*(z)$ is given by the  decay lengths 
  $1/\lambda_1,1/\lambda_2$ that differ from $\xi_0$ and 
  $\lambda_D^*$. Another important consequence
  of the above coupling between   $\Phi$ and $\rho_q^*$
  is the lack of a qualitative difference between the solutions of the linearized and nonlinear EL equations. 
 On the other hand, in our  theory for the inorganic salt qualitatively different results
   in the linearized and nonlinear 
   theories have been obtained~\cite{pousaneh:11:0,pousaneh:14:0}.
   
   We should note that our theory is similar to the theory developed by Onuki and Kitamura~\cite{onuki:04:0} and 
   studied in Ref.\cite{okamoto:11:0}. 
   The main difference between the two theories is the OP of the phase separation. In Refs.\cite{onuki:04:0,okamoto:11:0}
   the OP is identified with the concentration in the binary mixture, $\rho_w^*-\rho_l^*$, while we define the OP 
   in Eq.(\ref{OP}). We take into account the van der Waals interactions between the ions, and consider the 
   entropy of the four-component mixture, instead of a sum of the entropies of the two 2-component subsystems.
   Important  advantage of our approach is the link between 
   the mesoscopic and the microscopic description that was a starting point of our derivation. 
   Thanks to this link our theory is justified on a more fundamental level. 
    However,  in the case of the antagonistic salt 
   the difference between our theory and the theory of Onuki and Kitamura is less significant than
   in the case of the inorganic salt. In the latter case the nonlinear coupling between $\Phi$ and $\rho_q^*$
   plays a crucial 
   role in our theory, and is ignored in Ref.\cite{onuki:04:0,okamoto:11:0}. 
   
\begin{figure}[t]
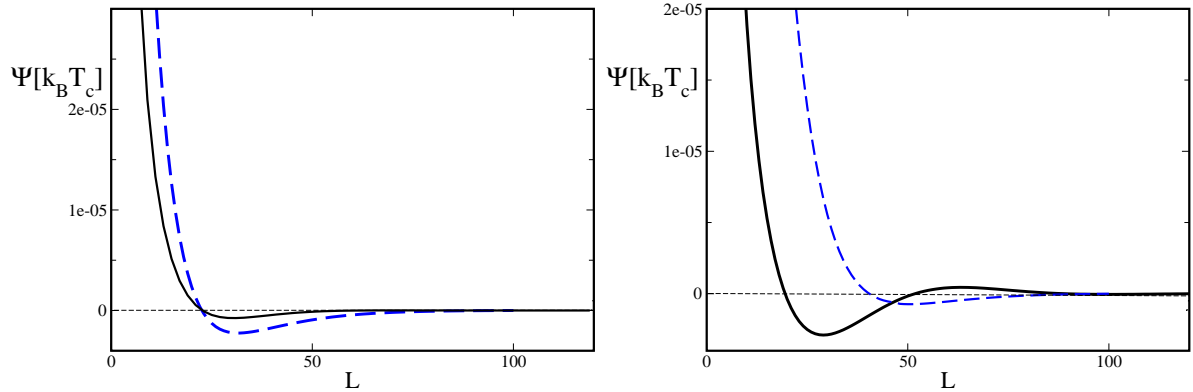

\hspace{-1.50cm}
 \includegraphics[scale=0.3]{./Figures/Fig10_a.eps}
  \includegraphics[scale=0.3]{./Figures/Fig10_b.eps}
\caption{ Effective potential obtained from numerical solutions of the full EL equations.  
Solid lines correspond to the current theory
while the  dash lines show the results of a theory of hydrophilic ions~\cite{pousaneh:12:0,pousaneh:14:0}.  The 
parameters are  $h=0.001$,  $\sigma=-0.001$, $\lambda_D^*=10$,  $\bar \rho_{ion}^* = 0.005$, (left)  $| T/T_c-1|= 0.005$, 
(right) $| T/T_c-1|= 0.002$. $L$ is in units of the microscopic distance $a=0.4 nm$.
}
\label{pot.11}
\end{figure}

   The effect of very  small amount of solute should be independent of its kind. Our functionals
   for the inorganic and antagonistic salt, however, differ significantly from each other. Nevertheless, we have verified
   that for small amount of ions and not very close to the critical point (for example for $\rho_{ion}^*=0.001$ 
   and $|T/T_c-1|=0.005$)
   the effective potential between parallel external surfaces has essentially the same form for the inorganic 
   and antagonistic salt.
   For larger $\rho_{ion}^*$ a quantitative difference between  $\Psi(L)$  in the presence of
   the inorganic  or the antagonistic salt can
   be seen (Fig.\ref{pot.11}). Qualitatively different shapes of $\Psi(L)$ 
   are obtained if  the correlation length is sufficiently 
   large compared to the period of the concentration oscillations. 
   The same amount of antagonistic salt leads to a deeper minimum of the effective potential at
   shorter separation than in the case of the inorganic salt, and the repulsion barrier occurs.  
   This shape of the potential may occur when
   $\rho_{ion}^*$ is big enough, and
   the critical point is approached (Fig.\ref{pot.11}).

   We have obtained a mesoscopic functional  of the
   Landau-Brazovskii form (\ref{LB}) from a microscopic density-functional theory for a four-component mixture
   by   a systematic coarse-graining procedure. A  functional of the same form 
   was successfully applied to amphiphilic systems
   \cite{gompper:94:0,ciach:01:2}. Thus,
   our  theoretical result and the experimental observations  
   of similarity between the mesoscopic structure
   induced by amphiphiles and antagonistic salt \cite{sadakane:13:0} are complementary.
   We have limited to the Gaussian approximation here. It is well known that beyond the Gaussian approximation 
   the fluctuations with the wavelength $k\simeq k_0$ yield a significant contribution to the correlation function,
   especially close to $t_0=0$. The $k$-dependence, however, is not changed up to a small correction; 
   only $t_0$ is renormalized in the 1-loop approximation. We fitted our 
    expression for $\tilde G(k)$ to the experimental curves. Since the 
   decay of the OP and the shape of the effective potential are determined by the correlation function, in our 
   results for $\Psi$ the renormalization of the model parameters is partially taken into account. 
   This way we avoid the overestimation
   of the mesoscopic structure typical for the men-field results.

   We have calculated the effective potential for 4 samples investigated experimentally
   in Ref.\cite{leys:13:0,sadakane:13:0}.
   The maximum of the structure factor 
   occurs for $k_0>0$ for three of these samples. 
   The oscillatory  concentration and effective potential can be seen only
   in the sample with the largest concentration of ions, however. 
   This is because the period of oscillations, $2\pi/\lambda_{im}$  is large compared
   to the decay length $1/\lambda_{re}$
   (Table I), whereas  a  detectable oscillatory potential can be expected
   for $1/\lambda_{re}>2\pi/\lambda_{im}$. 
   Based on our results (Fig.\ref{OPSadakane}) and on the analogy with surfactant mixtures where 
   such forces were measured~\cite{antelmi:99:0}, 
   we may expect oscillatory effective potential between the 
   confining surfaces  closer to the critical point or near 
   the transition to the lamellar phase.
   
   The potential in Figs.\ref{OPSadakane} and \ref{pot.11} is attractive at short separations 
   and repulsive at larger separations (SALR).
   For the SALR-type potentials between colloid particles finite clusters are expected, because the barrier prevents
   the clusters from further growth~\cite{stradner:04:0,ciach:13:03}. 
   
   We conclude that effective interactions between weakly charged colloid particles immersed
   in a near-critical mixture with antagonistic salt can exhibit very rich behavior. As a result, the particles can form
    different types of structures. Small changes of the amount of ions or temperature can lead 
    to qualitative changes of the interactions between the particles. 
   The sensitivity of the effective interactions to the thermodynamic state is a 
   property that may allow for
   manipulating with the structure of  colloids.

\acknowledgments

We thank Dr. Koichiro Sadakane and Dr. Jan Leys for providing us the data with their  experimental structure factor.  
The work  of FP  was realized within the International PhD Projects
Programme of the Foundation for Polish Science, cofinanced from
European Regional Development Fund within
Innovative Economy Operational Programme "Grants for innovation". AC  acknowledges
the financial support by the NCN grant 2012/05/B/ST3/03302.

\section{Appendices}
\subsection{
The Euler-Lagrange equations}

From 
 $\delta \Omega(\Phi({\bf r}),\rho_q^*({\bf r}),\rho_{ion}({\bf r}))/\delta {\rho_{ion}^*({\bf r})}=0$ we obtain 
  \begin{equation}
\label{mu2}
  \mu_{ion}=\frac{T^*}{2}\ln\Bigg(
\frac{\rho_{ion}^{*2}({\bf r})-\rho_q^{*2}({\bf r})}
{(1-\rho_{ion}^*({\bf r}))^2-(\Phi({\bf r})-\rho_q^*({\bf r}))^2}
\Bigg) .
  \end{equation}
In a uniform fluid Eq.(\ref{mu2}) takes the form
\begin{equation}
\label{mu1}
  \mu_{ion}=T^*\ln R,
 \end{equation}
where
 \begin{equation}
  R^2=\frac{\bar\rho_{ion}^{*2}}{(1-\bar\rho_{ion})^{*2} -\bar\Phi^2}.
 \end{equation}
By equating RHS of Eqs.~(\ref{mu1}) and (\ref{mu2}) we obtain 
  \begin{equation}
\label{rhoelim}
   \rho_{ion}^*({\bf r})=\frac{-R^2+\sqrt{R^2-(1-R^2)\Big[ R^2(\Phi({\bf r})-\rho_q^*({\bf r}))^2-\rho_q^*({\bf r})^2\Big]  }}
{1-R^2}.
  \end{equation}
With the help of Eq.~(\ref{rhoelim}) we can eliminate $\rho_{ion}^*({\bf r})$ from Eq.~(\ref{S}).
The remaining EL equations are obtained in a similar way, and have the forms
\begin{equation}
\label{epsi}
 e\beta\psi_{el}({\bf r})+\frac{1}{2}\ln\Bigg[
\frac{\Big(\rho_{ion}^*({\bf r})+\rho_q^*({\bf r})\Big)\Big(1-\Phi({\bf r})\Big)  + 
\Big(\rho_q^*({\bf r})^2 -\rho_{ion}^*({\bf r})^2\Big)}{\Big(\rho_{ion}^*({\bf r})-
\rho_q^*({\bf r})\Big)\Big(1+\Phi({\bf r})\Big)   + \Big(\rho_q^*({\bf r})^2 -\rho_{ion}^*({\bf r})^2\Big)}
\Bigg]=0
\end{equation}
and
\begin{equation}
\label{EL3}
\frac{d^2\Phi({\bf r})}{d {\bf r}^2}= -b{\Phi({\bf r})}+\frac{T^*}{2}\ln\Bigg[
\frac{1-\rho_{ion}^*({\bf r})+\Phi({\bf r})-\rho_q^*({\bf r})}{1-\rho_{ion}^*({\bf r})-\Phi({\bf r})+\rho_q^*({\bf r})}
\Bigg].
\end{equation}
Eqs.~(\ref{rhoelim})-(\ref{EL3}) and (\ref{Poisson}) form a closed set of two differential
 and two algebraic equations. 
 
 When
  $\rho_q^*({\bf r})$, $\varPhi$
and
\begin{eqnarray}
 \vartheta({\bf r})=\rho_{ion}^*({\bf r})-\bar\rho_{ion}^*
\end{eqnarray}
 are small, then we can solve analytically the linearized EL equations. Here we focus on the critical composition, 
  $\bar\Phi=0$. 
 Let us limit ourselves to the functions that depend only on $z$. From (\ref{EL3}) we obtain 
 \begin{equation}
\label{EL1}
\frac{d^2\Phi(z)}{dz^2}= \xi_0^{-2} \Phi(z)   -\frac{T^*}{(1-\bar \rho_{ion}^*)}\rho_q^*(z),
\end{equation}
and from (\ref{epsi}) and  (\ref{Poisson}) we have
 \begin{equation}
\label{EL2}  
\frac{d^2\rho_q^*(z)}{dz^2}=\kappa^2(1- \bar\rho_{ion}^*)\rho_q^*(z)+\bar \rho_{ion}^* \frac{d^2\Phi(z)}{dz^2}.
\end{equation}

\subsection{
Internal energy in Fourier representation}

The electrostatic energy Eq.~(\ref{el}) in the Fourier representation is given by
\begin{equation}
\label{elfou}
  U_{el}=\frac{1}{2}\int_V \frac{d{\bf k}}{(2\pi)^d} \bigg(\frac{4\pi e^2}{a\epsilon k^2}\bigg )
  \tilde \rho_q^*({\bf k})\tilde \rho_q^*({\bf k}),\\
 \end{equation} 
where the wavenumber $k$ is dimensionless (in $a^{-1}$ -units) and we used the Poisson equation
in Fourier representation
\begin{equation}
\label{elfouk}
 k^2 \tilde \psi_{el}({\bf k}) =\frac{4\pi e}{a\epsilon} \tilde \rho_q^* ({\bf k}).
 \end{equation} 
The van der Waals contribution Eq.~(\ref{uvdW}) in the  Fourier representation  takes the form 
\begin{equation}
\label{vWfou}
 U_{vdW}=\frac{J}{2}\int_V \frac{d{\bf k}}{(2\pi)^d}   (-b+k^2) \tilde \Phi({\bf k}) \tilde \Phi(-{\bf k}).\\
 \end{equation} 
 
\subsection{
Parameters in the solutions of the EL equations}

The amplitudes in Eq.(\ref{Phi(z)}) are
\begin{equation}
 \left\{ 
  \begin{array}{l }
   A_1=\frac{h_0(\lambda_2^2-\zeta^{-2})\lambda_2-
 \frac{T^*\sigma_0}{\lambda_D^{*2}}(1+\lambda_2)}{D}, 
\vspace{0.4cm}\\
  A_2=-\frac{h_0(\lambda_1^2-\zeta^{-2})\lambda_1-\frac{T^*\sigma_0}{\lambda_D^{*2}}(1+\lambda_1)}{D}.
  \end{array} \right.
\end{equation}
with
\begin{equation}
 D=(\lambda_1-\lambda_2)[\zeta^{-2}-\lambda_1^2-\lambda_2^2-\lambda_1\lambda_2(1+\lambda_1+\lambda_2)]
\end{equation}

where  $\lambda_i$ and $\zeta$ are given  in Eq.(\ref{delta12})  and  (\ref{zeta}) respectively.
%
The real and imaginary parts of the inverse decay lengths $\lambda_i$ are
\begin{equation}
\label{lambdare}
 \left\{ 
  \begin{array}{l }
     \lambda_{re}^2=\frac{1}{4\lambda_D^{*2}}\Bigg[
 \Big(
 \sqrt{a_D} +\frac{\lambda_D^*}{\xi_0}
 \Big)^2-a_N
 \Bigg], 
\vspace{0.4cm}\\
  \lambda_{im}^2=\frac{1}{4\lambda_D^{*2}}\Bigg[a_N-
 \Big(\frac{\lambda_D^*}{\xi_0}-
 \sqrt{a_D} 
 \Big)^2
 \Bigg] .
  \end{array} \right.
\end{equation}

\end{document}